\documentclass[a4paper,usenatbib]{mem}
\usepackage{natbib}
\usepackage{graphicx}
\usepackage[a4paper]{hyperref}

\newcommand{\be}{\begin{equation}}
\newcommand{\ee}{\end{equation}}
\newcommand{\ba}{\begin{eqnarray}}
\newcommand{\ea}{\end{eqnarray}}
\newcommand{\brr}{\begin{array}}
\newcommand{\err}{\end{array}}
\newcommand{\bc}{\begin{center}}
\newcommand{\ec}{\end{center}}

\newcommand{\msun}{\,h^{-1}M_\odot}

\newcommand{\lum}{\,{\rm erg\,s^{-1}}}

\newcommand{\fnl}{f_\mathrm{NL}}

\newcommand{\mincir}{\raise
  -2.truept\hbox{\rlap{\hbox{$\sim$}}\raise5.truept \hbox{$<$}\ }}
\newcommand{\magcir}{\raise
  -2.truept\hbox{\rlap{\hbox{$\sim$}}\raise5.truept \hbox{$>$}\ }}
\newcommand{\siml}{\raise
  -2.truept\hbox{\rlap{\hbox{$\sim$}}\raise5.truept \hbox{$<$}\ }}
\newcommand{\simg}{\raise
  -2.truept\hbox{\rlap{\hbox{$\sim$}}\raise5.truept \hbox{$>$}\ }}

\idline{73}{23}

\begin{document}
   \title{Astrophysics and cosmology with galaxy clusters:
the WFXT perspective
}

   \author{S. Borgani\inst{1,2,3}, P. Rosati\inst{4}, B. Sartoris\inst{1,2,3},  
   P. Tozzi\inst{2,3}, R. Giacconi\inst{5}, \& the WFXT Team\\
}

\institute{Dipartimento di Fisica, Sezione di Astronomia, Universit\`a
  di Trieste, Via Tiepolo 11, I-34143 Trieste, Italy
  \email{borgani,sartoris@oats.inaf.it}
  \and INAF-Osservatorio Astronomico di Trieste, Via Tiepolo 11,
  I-34143
  Trieste, Italy \email{tozzi@oats.inaf.it}
  \and
  INFN, Sezione di Trieste, Via Valerio 2, I-34127 Trieste, Italy
  \and 
  ESO-European Southern Observatory, D-85748 Garching bei
  M\"unchen, Germany \email{prosati@eso.org}
  \and
  Department of Physics and Astronomy, The Johns Hopkins University,
  Baltimore MD, USA
} 

\abstract{We discuss the central role played by
  the X-ray study of hot baryons within galaxy clusters to reconstruct
  the assembly of cosmic structures and to trace the past history of
  star formation and accretion onto supermassive Black Holes (BHs). We
  shortly review the progress in this field contributed by the current
  generation of X-ray telescopes. Then, we focus on the outstanding
  scientific questions that have been opened by observations carried
  out in the last years and that represent the legacy of Chandra and
  XMM: (a) When and how is entropy injected into the inter-galactic
  medium (IGM)? (b) What is the history of metal enrichment of the
  IGM? (c) What physical mechanisms determine the presence of cool
  cores in galaxy clusters? (d) How is the appearance of
  proto-clusters at z~2 related to the peak of star formation activity
  and BH accretion? (e) What do galaxy clusters tell us about the
  nature of primordial density perturbations and on the history of
  their growth? We show that the most efficient observational
  strategy to address these questions is to carry out a large-area
  X-ray survey, reaching a sensitivity comparable to that of deep
  Chandra and XMM pointings, but extending over several thousands of
  square degrees. A similar survey can only be carried out with a
  Wide-Field X-ray Telescope (WFXT), which combines a high survey
  speed with a sharp PSF across the entire FoV. We emphasize the
  important synergies that WFXT will have with a number of future
  ground-based and space telescopes, covering from the radio to the
  X-ray bands. Finally, we discuss the immense legacy value that such
  a mission will have for extragalactic astronomy at large.

   }
\authorrunning{S. Borgani et al.}  \titlerunning{Galaxy clusters with
  WFXT}
   \maketitle
%

\section{Introduction}
Galaxy clusters represent the place where astrophysics and cosmology
meet: while their overall internal dynamics is dominated by gravity,
the astrophysical processes taking place on galactic scale leave
observable imprints on the diffuse hot gas trapped within their
potential wells
\citep{rosati02,voit05,borgani_kravtsov09}. Understanding in detail
the relative role played by different astrophysical phenomena in
determining this cosmic cycle of baryons, and its relationship with
the process of galaxy formation, is one of the most important
challenges of modern cosmology.  Clusters of galaxies represent the
end result of the collapse of density fluctuations over comoving
scales of about 10 Mpc. For this reason, they mark the transition
between two distinct dynamical regimes. On scales roughly above 10
Mpc, the evolution of the structure of the universe is mainly driven
by gravity and the evolution still feels the imprint of the
cosmological initial conditions. At scales below 1 Mpc the physics of
baryons plays an important role in addition to gravity, thus making
physical modeling far more complex. In the current paradigm of
structure formation, clusters form via a hierarchical sequence of
gravitational mergers and accretion of smaller systems. Within these
small halos gas efficiently cools, forms stars and accretes onto
supermassive black holes (SMBHs), living in massive galaxies, already
at high redshift. While the star formation peaks at redshift $z\sim
2$--3, the intergalactic gas is heated to high, X-ray emitting
temperatures by adiabatic compression and shocks, and settles in
hydrostatic equilibrium within the cluster potential well, only at
relatively low redshift, $z\mincir 2$. The process of cooling and
formation of stars and SMBHs can then result in energetic feedback due
to supernovae or AGN, which inject substantial amounts of heat into
the intergalactic medium (IGM) and spread heavy elements throughout
the forming clusters.

Galaxy clusters are also very powerful cosmological tools. They probe
the high end of the mass function of dark matter (DM) halos, whose
evolution is highly sensitive to the underlying cosmological scenario
and to the growth rate of cosmological perturbations
\citep[e.g.,][]{borgani01,voit05}. This information, combined with the
shape and amplitude of the power spectrum of their large-scale
distribution, offers a means of constraining the growth of cosmic
structures over a wide range of scales. For these reasons, galaxy
clusters are nowdays considered sensitive probes of the dark sector of
the Universe and of the nature of gravity, complementary to CMB and
SN-Ia tests, which are sensitive to the backgorund geometry and
expansion rate. Based on relatively small samples of few tens of distant
X--ray clusters extracted from ROSAT deep pointings, followed up by
Chandra observations, independent analyses have recently shown that
the evolution of the population of galaxy clusters does indeed provide
significant constraints on cosmological parameters
\citep[e.g.][]{vikhlinin09c,mantz09I}. This remarkable progress in
cluster cosmology has been made possible by the introduction of robust
X--ray mass proxies, such as the gas mass $M_{gas}$ and the total
thermal content of the ICM defined by the product of gas mass and
temperature, $Y_X=M_{gas}T$ \citep[e.g.][]{kravtsov06}. Quite
interestingly, the scatter in the relation between such mass proxies
and the total cluster mass is suppressed after excising core cluster
regions, $\mincir 0.15 R_{500}$.

Such results demonstrate that, to fully exploit the potential of
clusters for cosmological applications, detailed measurements of
X--ray mass requires collecting an adequate number of photons and good
spatial resolution to remove the contribution of core regions in
distant objects.  From one hand, the revitalization of cluster
cosmology has indeed required the high data quality offered by the
present generation of X-ray satellites.  On the other hand, it
highlights the constraining power that future X--ray surveys, like the
one to be provided WFXT, could provide. The WFXT surveys would
increase by several orders of magnitude the statistics of distant
clusters for which data of comparable quality as that provided by
Chandra observation.

As we will discuss in this contribution, the large grasp of WFXT
combined with its sharp and stable PSF makes it the ideal instrument
for astrophysical and cosmological studies of galaxy clusters (see
also \citealt{giacconi09}, \citealt{vikhlinin09b}, and
Rosati et al., in this volume).

\section{WFXT to study clusters as astrophysical laboratories}
Thanks to the high density and temperature reached by the gas within
their potential wells, galaxy clusters mark the only regions where
thermo- and chemo-dynamical properties of the IGM can be studied in
detail at $z<1$ from X--ray emission, and directly connected to the
optical/near-IR properties of the galaxy population. A remarkable leap
forward in the quality of X-ray observations of clusters took place
with the advent of the Chandra and XMM-Newton satellites. Thanks to
their unprecedented sensitivity (and angular resolution in case of
Chandra), they led to a number of fundamental discoveries concerning
nearby, $z\mincir 0.3$, clusters. For instance:
\begin{description}
\item[(a)] The lack of strong emission lines at soft X-ray energies in the
core regions placed strong limits on the amount of gas that can cool
to low temperatures \citep{peterson06}, thus challenging the classical “cooling flow”
model \citep{fabian94};
\item[(b)] Temperature profiles have been unambiguously observed to
  decline outside the core regions and out to the largest radii
  sampled so far, $\sim R_{500}$\footnote{We indicate with $R_\Delta$
    the cluster-centric radius encompassing an average overdensity $\Delta$
    times the critical cosmic density. For reference, $\Delta=200$ is
    close to the virial overdensity while $\Delta=500$ corresponds to
    about half the virial radius for a concordance $\Lambda$CDM
    model.}, while they gently decline toward the cluster center in
  relaxed systems\citep[e.g.][]{vikhlinin05,pratt07,leccardi08a};
\item[(c)] The level of gas entropy at $R_{500}$ is in excess of what
  explainable by the action of supersonic accretion shocks
  \citep[e.g.][]{sun09,pratt09}, while it is unexpectedly low in the
  innermost regions of relaxed clusters \citep[e.g.][]{donahue06};
\item[(d)] The intra-cluster medium (ICM) is not uniformly enriched in
metals, instead metallicity profiles are observed to have a spike in
the central regions, associated to the presence of the brightest
cluster galaxy (BCG), while declining at least out to $\simeq 0.3R_{500}$.
\end{description}

While these observations shed new light on our understanding of the
physical properties of the low-redshift intergalactic medium, (IGM),
they opened at the same time at least as many questions.
As we will discuss here below, an efficient way of addressing open
questions in the ICM study is by carrying out high--sensitivity X-ray
surveys, which provide a large number of clusters for which detailed
studies can be carried out at low and high redshift. 
In Figure 1, we show a comparison for the yields of clusters expected
from five years of operation of WFXT, compared with the expectations
for the planned German-led mission eROSITA\footnote{Based on the
  specifications as provided in Mission Definition Document ({\tt
    http://www.mpe.mpg.de/erosita/MDD-6.pdf})}. Besides the huge
number of clusters that WFXT will detect at large redshift, this
demonstrates that measurements of the physical properties of the ICM
will be available for a large number of them.

   \begin{figure}
   \centering
   \includegraphics[width=7.truecm]{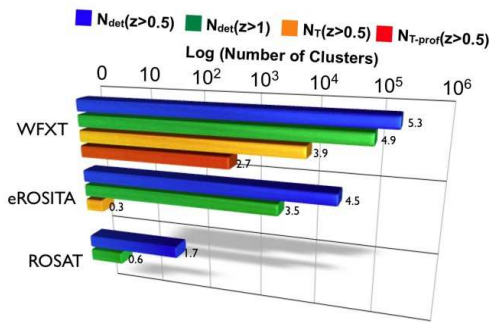}
   \caption{The comparison between numbers of clusters expected for
     the surveys to be carried out by WFXT (as described in the
     contribution by Rosati et al. in this volume), by eROSITA, and as
     found in the ROSAT All Sky Survey \protect\citep{RASS}. Blue and
     grees bars: number of clusters detected at $z>0.5$ and $z>1$,
     respectively; yellow and orange bars: number of clusters with at
     least $1.5\times 10^3$ and $1.5\times 10^4$ net photon counts in
     the [0.5-2] keV energy band, respectively.}
         \label{counts}
   \end{figure}


\noindent {\bf When and how is entropy injected into the
IGM?} The standard explanation for the excess of entropy observed
out to $R_{500}$ is that some energetic phenomena should have
heated the ICM over the cluster life-time
\citep[e.g.][]{voit05,borgani_kravtsov09}. Models based on the
so-called pre-heating (i.e. diffuse entropy injection before the
bulk of the mass is accreted into the cluster halos) have been
proposed as an explanation
\citep[e.g.][]{tozzi01,borgani02,voit03}. However, these models
predict quite large isentropic cores, which are not
observed. Furthermore, studies of the intergalactic medium
(IGM), from observations of $z\magcir 2$ absorption systems in
high-resolution optical spectra of distant QSOs, demonstrate that
any pre-heating should take place only in highdensity regions
\citep{shang07,borgani_viel09}. An alternative scenario is that ICM
heating takes place at relatively low redshift, within an already
assembled deep potential well. In this case, the natural
expectation is that the same heating agent, presumably the central
AGN, should be responsible for both establishing the cool core and
increasing the entropy out to $\sim 1$ Mpc scale, although it is
not clear how AGN feedback can be distributed within such a large
portion of the cluster volume.

Reconstructing the timing and pattern of entropy injection in the ICM
has far reaching implications for tracing the past history of star
formation and black hole (BH) accretion. While we expect that the two
above scenarios leave distinct signatures on the evolution of the ICM
entropy structure, available data from XMM and Chandra are too sparse
to adequately understand this evolution.

The large number of clusters with $\sim 10^4$ counts available from the
WFXT surveys would increase by orders of magnitude the statistics of a
handful of clusters at $z>0.5$ for which detailed entropy profiles
have been measured so far. The measurement of ICM profiles for a large
number of distant
clusters will allow us to trace the interplay between IGM and galaxy
population along 2/3 of the cosmological past light-cone. Furthermore,
the low background and the possibility of resolving out the
contribution of point sources will also allow us to measure such
profiles out to $R_{200}$ and beyond for bright galaxy clusters at
$z<0.2$ (see the contribution by Ettori \& Molendi, in this volume).

\noindent {\bf What is the history of metal enrichment of the IGM?}
This question is inextricably linked to the previous one on the
history of IGM heating. Measurements of the metal content of the ICM
provide direct information on the past history of star formation and
on processes (e.g., galactic ejecta powered by SN and AGN,
ram-pressure stripping of merging galaxies, stochastic gas motions,
etc.) that are expected to displace metal-enriched gas from star
forming regions \citep[e.g.,][]{schindler08}. So far, X--ray
observations have provided valuable information on the pattern of
enrichment only at low-redshift, $z\mincir 0.3$
\citep[][]{baumgartner05,mushotzky04,werner06}. Profiles of the Fe
abundance have been measured for nearby systems
\citep[e.g.][]{snowden08,leccardi08b}.  However, these results are
limited to rather small radii, $\mincir 0.3R_{500}$, while the level
of enrichment at larger radii should be quite sensitive to the timing
of metal production and to the mechanism of metal
transport. Furthermore, profiles of chemical abundances for other
elements, such as O, Si, and Mg, are much more uncertain. Tracing the
relative abundances of different chemical species, which are
synthetized in different proportions by different stellar populations,
is crucial to infer the relative role played by different SN types and
to establish the time-scale over which the ICM enrichment took
place. The situation is even more uncertain at $z>0.3$. Although
analyses based (mainly) on the Chandra archive show indications for an
increase of the ICM enrichment since $z\sim 1$
\citep{balestra07,maughan08}, basically no information is available on
the metallicity profiles and on abundance of elements other than
Fe. To improve upon this situation, one needs (a) to push to larger
radii the study of the distribution in the ICM of different chemical
species in nearby clusters; (b) to measure profiles of the Fe
abundance for hundreds of clusters at $z>0.5$.

Iron metallicity profiles would be measured by WFXT for virtually all
the clusters for which a temperature profile is obtained, although
with $\sim 2$ times larger statistical errors. A very accurate
measurement of the global Fe metallicity will be obtained for several
thousands of clusters out to $z\sim 1.5$. For all the clusters of this
sample, thermo-dynamical and chemical properties of the ICM will be
characterized with unprecedented precision.

\noindent {\bf What physical mechanisms determine the presence of cool
  cores in galaxy clusters?} XMM and Chandra unambiguously
demonstrated that the rate of gas cooling in cluster cores is
unexpectedly low. Such a low cooling rate requires that some sort of
energy feedback must heat the ICM so as to exactly balance radiative
losses. AGN are generally considered as the natural solution to this
problem \citep[e.g.][]{mcnamara07}. However, no consensus has been
reached so far on the relative role played by AGN and by mergers in
determining the occurrence of cool cores in galaxy clusters
\citep[e.g.][]{burns08}. Since merging activity and galactic nuclear
activity are both expected to evolve with redshift, measurements of
the occurrence of cool cores in distant clusters are necessary to
address this issue. Although attempts have been pursued to
characterize the evolution of the fraction of cool cores using Chandra
data \citep[e.g.][]{Santos08}, no definite conclusion has been reached
on the evolution of the fraction of cool core clusters.

The sharp and stable PSF of WFXT will allow one to resolve
the core region of distant clusters (a cool-core of 50 kpc radius will
subtend an angle of $\simeq 6$ arcsec at $z=1$). The yield of hundreds of
clusters at $z>0.5$ for which more than $10^4$ counts will be available,
will allow us to accurately measure the evolution of the occurrence of
cool cores and how this is related to the cluster dynamical state.

\noindent {\bf How is the appearance of proto-clusters related to the
  peak of star formation activity and BH accretion?}
Massive galaxies in today’s clusters show only very modest ongoing
star formation: they harbor a super-massive black hole usually living
in a quiescent accretion mode and experience only ``dry'' mergers with
much smaller galaxies. The situation should be radically different at
$z\sim 2$. This is the epoch when proto-BCGs are expected to be
assembling through violent mergers between actively star-bursting
galaxies, moving within a rapidly evolving potential well. These
proto-cluster regions accrete a large amount of gas that is suddenly
heated to high temperature by mechanical shocks and, for the first
time, starts radiating in X-rays. At the same time, BHs hosted within
merging galaxies are expected to coalesce and their accretion disks to
be destabilized by the intense dynamical activity, thereby triggering
a powerful release of feedback energy. Evidence for such forming
proto-clusters has been obtained by optical observations of a strong
galaxy overdensity region, the so-called Spiderweb complex,
surrounding a previously identified powerful radio galaxy at $z\simeq
2.1$ \citep{miley06,hatch09}. Cosmological simulations lend support to
the expectation that similar structures trace the progenitors of
massive cluster seen locally, and predict that this structure should
already contain dense IGM, emitting in X-rays with $L_X\sim 10^{44}
\lum$ in the [0.5-2] keV band, with a temperature of several keV and
enriched in metal at a level comparable to nearby clusters
\citep{saro09}. As of today, no unambiguous detection of X-ray
emitting gas permeating this region has been obtained
\citep{carilli02}. While the detection of such a hot diffuse gas may
be just at the limit of the capability of current X-ray telescopes,
characterizing its physical properties (temperature and metallicity)
is far beyond the reach of Chandra and XMM. 

The study of proto-clusters at $z\magcir 2$ is still unexplored
territory. For this reason, it is difficult to make predictions on how
many of these structures could be observed by WFXT. By extrapolating our
present knowledge of the relation between mass and X-ray luminosity,
we expect to detect several hundreds of such objects over the whole
sky. For the brightest of these clusters, it will even be possible to
measure their redshift through X-ray spectroscopy with deeper
follow-up exposures. At $z\sim 2$ the inverse Compton scattering of
relativistic electrons, injected by AGN in core regions, off the CMB
photons is much more effective than at low-$z$ in producing a hard
X-ray excess, thanks to the higher CMB temperature. Based on the
expectation from hydrodynamic simulations, we estimated that 5 to 10
thousands of photons would be detectable by WFXT in a deep 400 ksec
pointing on a $z\simeq 2$ proto-cluster, which is the progenotor of a
today massive cluster, with $M_{200}\simeq 10^{15}\msun$. Such an
observation would allow one: (a) to catch ``in fieri'' the pristine ICM
enrichment; (b) to see in action the combined effect of strong mergers
and intense nuclear activity within a forming cluster; (c) to discern
the thermal and non-thermal emission from X-ray spectroscopy and infer
the early contribution of cosmic rays in pressurizing the ICM.

\vspace{0.3truecm} The goal of measuring physical properties of the
ICM out to $z\sim 1$ and beyond can only be accomplished by a survey
with the area and sensitivity achievable with WFXT. In fact, WFXT
constitutes a two orders of magnitude improvement with respect to
eROSITA (similar to the area-sensitivity enhancement that eROSITA will
give with respect to the ROSAT All-Sky Survey), with in addition a 5
times better angular resolution (see the contribution by Cappelluti et
al, in this volume). 

\section{Cluster cosmology with WFXT}
WFXT will not be just a highly efficient cluster-counting machine. Its
unique added value is that it will characterize the physical
properties for a good fraction of these clusters and, therefore,
calibrate them as robust tools for cosmological applications. Based on
the specification of the WFXT surveys (see Rosati et al., this
volume), we computed the constraints that can be placed on different
classes of cosmological models. By following the approach described by
\cite{sartoris10}, we apply the Fisher-Matrix technique to forecast
constraints on cosmology by combining information from number counts
and power spectrum of clusters. The computation of these forecasts is
based on the so-called self--calibration approach
\citep[e.g.,][]{majumdar03,lima04}. In this approach, we assume that
X--ray observations provide an estimate of the actual cluster mass
whose uncertain relation with the actual cluster mass is described by
a suitable set of '`nuisance'' parameters, to be fitted, with their
own priors, along with cosmological parameters.

\begin{figure*}
\centering
\hbox{
\includegraphics[width=0.45\textwidth,angle=270]{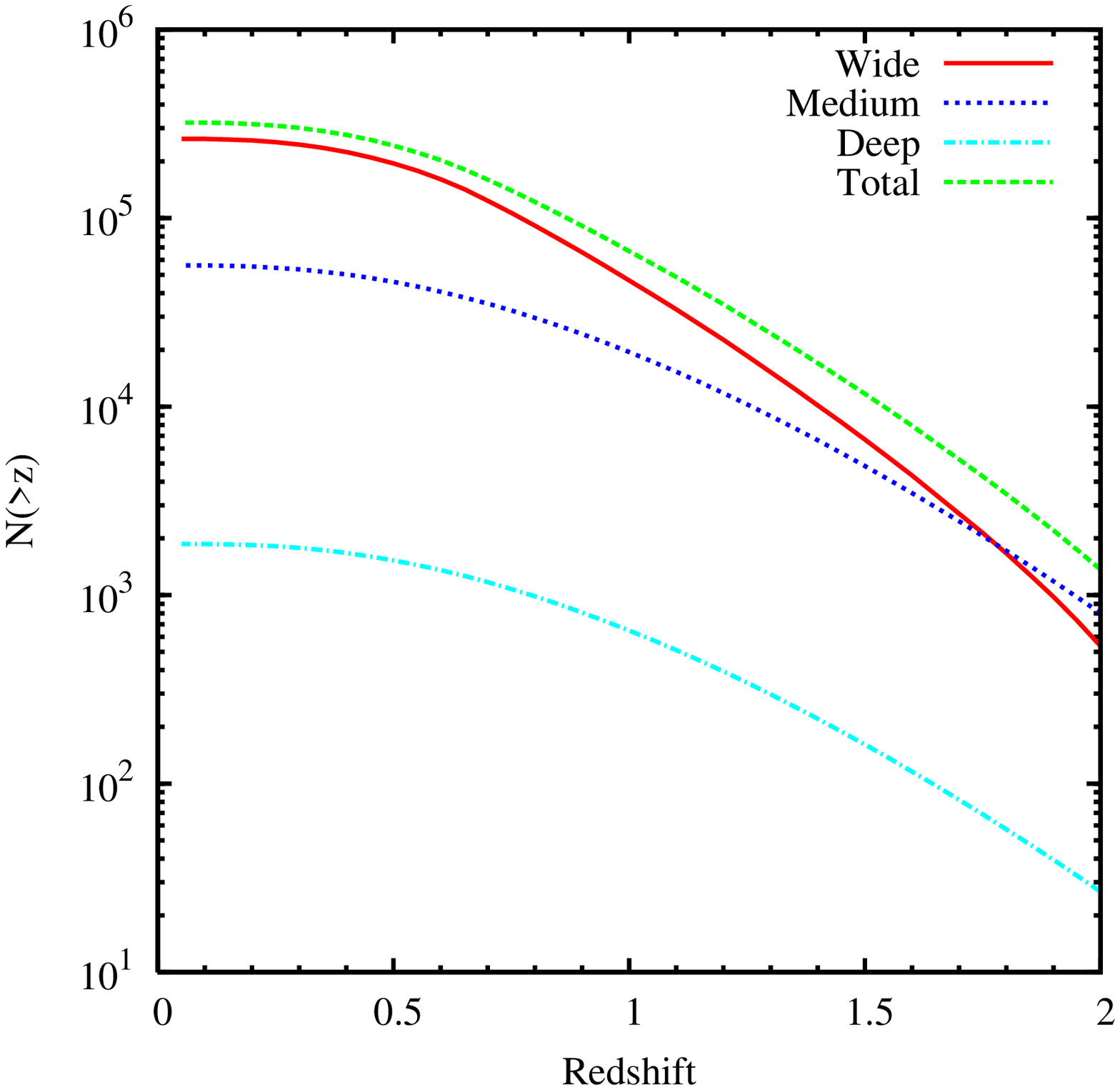}
\hspace{-2.5truecm}
\includegraphics[width=0.45\textwidth,angle=270]{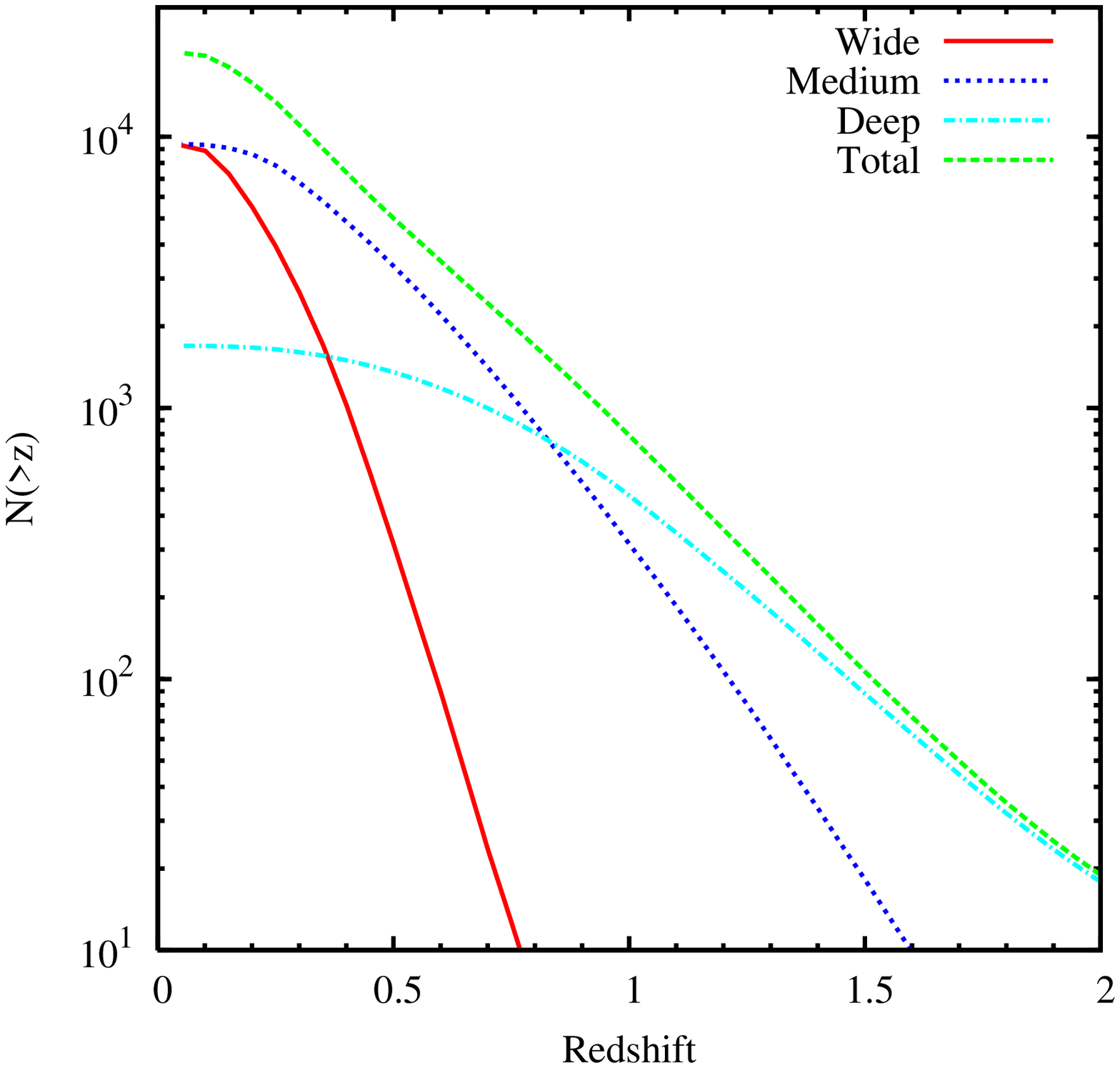}
}
\caption{The cumulative redshift distributions for the three WFXT
  surveys. The left panel is for all clusters detected, while the
  right panel is for the clusters in the ``Golden Samples'',
  corresponding to a 1500 photons detected. In both panels solid
  (red), dotted (blue) and dot-dashed (cyan) curves are for the Wide,
  Medium and Deep surveys, respectively, while the short-dashed
  (green) curve is for the sum of the three surveys.}
\label{nz}
\end{figure*}

\cite{sartoris10} used this approach to place constraints on possible
deviations from Gaussianity of the primordial density fluctuation
field. The reference cosmological model assumed in this analysis,
consistent with the WMAP-7 best--fitting model \citep{komatsu10},
assume: $\Omega_m=0.28$ $\sigma_8=0.81$, $\Omega_k=0$ for the
curvature, $w(a)=w_0+(1-a) \; w_a$ with $w_0=-0.99$ and $w_a=0$ for
the Dark Energy equation of state, $\Omega_{\rm b}=0.046$ for the
baryon contribution, $h=0.70$ for the Hubble parameter, $n=0.96$ for
the primordial spectral index and $\fnl=0$ for the non--Gaussianity
parameter. Furthermore, in the analysis we also include priors on
cosmological parameters as expected from the Planck CMB experiment
\citep{rassat08}. 

We adopt the appropriate flux-dependent sky coverages for the three
surveys (see Tozzi et al., this volume). To convert fluxes into
masses, we use the relation between X--ray luminosity and $M_{500}$
calibrated by \citet{maughan07}, where masses are recovered from
$Y_X$, using Chandra data for 115 clusters in the redshift range
$0.1<z<1.3$.  The relation between measured and true mass is described
by four nuisance parameters, which describe a possible intrinsic bias
in the mass estimate, e.g. related to a residual violation of
hydrostatic equilibrium \citep[e.g.][]{rasia06,piffaretti08,lau09}, an
intrinsic scatter in this relation and the evolution of these two
parameters \citep[see][ for a more detailed discussion]{sartoris10}.

In the left panel of Figure \ref{nz} we show the redshift distribution for
all the clusters detected in the three WFXT surveys, having mass of at
least $M_{500}>5\times 10^{13}\msun$. The right panel shows the same
redshift distributions for the ``Golden Samples'', i.e. for all the
clusters which are detected with at least 1500 net photon
counts. The left panel demonstrates the huge potential of WFXT to
detect a large number of clusters out to $z\sim 2$ and
beyond. Furthermore, the right panel demonstrates that WFXT is not
only a highly efficient survey machine to count clusters. In fact, its
large grasp also provides a large enough number of photons and,
therefore, to precise measurements of robust mass proxies, for about
20,000 clusters, with $\sim 1000$ of them at $z>1$. This represents a
huge improvement with respect to the few tens of distant clusters
available at present. This plot also shows the relevence of the Deep
Survey to calibrate measurements of mass proxies beyond $z\sim 1$,
thus complementing the larger statistics of lower-$z$ clusters
provided by the Medium and Wide surveys.

\begin{figure}
\hspace{-0.25truecm}
\includegraphics[width=0.50\textwidth,angle=270]{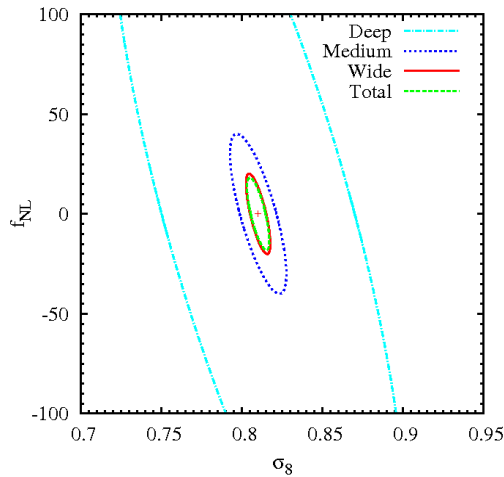}
\caption{Constraints at the 68 per cent confidence level on
  non--Gaussian parameter $\fnl$ and power spectrum normalization
  $\sigma_8$ from the Deep, Medium and Wide surveys (solid red,
  long-dashed blue and short-dashed black curves, respectively), by
  combining number counts and power spectrum information. Also shown
  with the dotted magenta curve are the constraints obtained from the
  combination of the three surveys. No prior knowledge is assumed for
  the values of the nuisance parameters. The Fisher Matrix from Planck
  experiment is included in the calculation of all constraints.}
\label{fig:det}
\end{figure}

We show in Figure \ref{fig:det} the joint constraints on the $f_{NL}$
parameter, which define the deviation from Gaussianity \citep[e.g.][
for recent reviews]{verde10,desjacques10} and the normalization of the
power spectrum, $\sigma_8$, after marginalizing over all the other
cosmological and nuisance parameters. As discussed by
\cite{sartoris10} \citep[see also][]{oguri09,cunha10}, constraints on
non-Gaussianity are weakly sensitive to the uncertain knowledge of the
nuisance parameters. On the other hand, non-Gaussian constraints
mainly comes from the shape of the power spectrum at the long
wavelengths probing the possible scale dependence of the biasing
parameter.  For these reasons, we used in this analysis the large sets
of detected clusters, without restricting to the ``Golden Sample'' for
which nuisance parameters can be measured. This plot clearly shows
that most of the constraints on non-Gaussianity comes form the Wide
survey, which in fact has the potential to prove the long wavelength
modes. Little information is carried instead by the Deep survey, which
is instead very important for the calibration of mass proxies for
distant clusters.

\begin{figure}
\hspace{-0.2truecm}
\includegraphics[width=0.47\textwidth,angle=270]{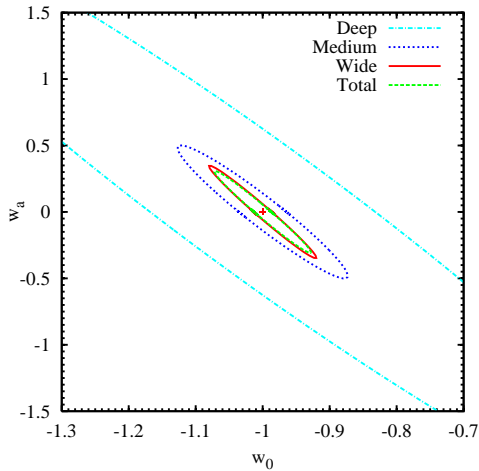}
\caption{The same as in Fig. \protect\ref{fig:det}, but for the
  constraints on the DE equation of state. The analysis is carried out
  for the same samples and same priors on nuisance parameters as in
  Fig.\protect\ref{fig:det}.}
\label{fig:DE}
\end{figure}

Figure \ref{fig:DE} shows the constraints on the parameters defining
the DE equation of state using the samples and the same priors on
nuisance parameters as for Fig.\ref{fig:det}. Also in this case, the
large cluster statistics available in the Wide survey, makes it
providing the dominant constraining power. The resulting value of the DETF
Figure-of-Merit \cite{albrecht09}, after combining the information
from the three surveys, is $DETF=512$.

\begin{figure}
\hspace{-1.75truecm}
\includegraphics[width=0.50\textwidth,angle=270]{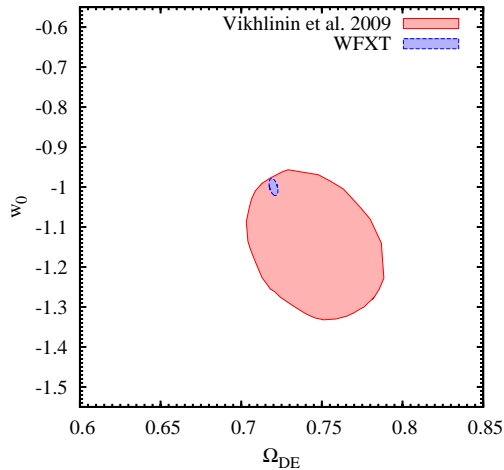}
\vspace{-0.45truecm}
\caption{A comparison between the constraints on the
  $\Omega_{DE}$--$w_0$ plane obtained by \protect\cite{vikhlinin09c}
  from a sample of clusters followed-up with Chandra (red shaded
  area), and as expected for the ``Bright'' WFXT surveys, by combining
  number counts and power spectrum information, under the assumption
  of strong strong priors on the nuisance parameters (light blue
  filled area). Both contours corresponds to $\Delta\chi^2 =1$
  (i.e. 68 per cent confidence level for one significant parameter)
  and are obtained under the assumption of flat Universe.}
\label{fig:DE_vik}
\end{figure}

Figure \ref{fig:DE_vik} emphasizes the improvement represented by WFXT
with respect to present constraints from X--ray cluster surveys.  In
this plot, the red shaded area show the constraints on the
$\Omega_{DE}$--$w_0$ plane obtained by \cite{vikhlinin09c} from a
sample combining nearby and distant cluster, originally identified
from ROSAT data and followed-up with Chandra. Since Chandra follow-up
provides at least $\sim 10^3$ photons per cluster, for consistency we
compare it with the forecasts for the WFXT samples (light blue
ellipse). The WFXT contour, which is obtained by combining number
counts and power spectrum information, is off-centered with respect to
the contours by \cite{vikhlinin09c} since their best--fitting model
does not coincide with the reference cosmological model assumed for
our Fisher--Matrix analysis of forecasts.

As well known, the evolution of the population of galaxy clusters is
affected by cosmology both through the cosmic expansion history, which
defines the volumes, and through the linear growth rate of
perturbations \citep[e.g.][]{haiman01}. In order to make a pure test
of perturbation growth, we decided to carry out the Fisher-Matrix
analysis by freezing the expansion history to the $\Lambda$CDM one,
while using a suitable parametrization to describe the growth
history. A commonly adopted approach to parametrize the growth of
perturbations is based on the quantity
\be
f(a)\,=\,{d\log D_{+}(a)\over d\log a}\,,
\label{eq:grw}
\ee
where $a$ is cosmic expansion factor and $D_+(a)$ is the linear growth
rate of density perturbations. The quantity $f(a)$ is well
approximated by the phenomenological relation
\citep[e.g.][]{wang_steinhardt98}
\be
f(a)\simeq \Omega_m(a)^\gamma
\label{eq:gam}
\ee
with $\gamma=0.55+0.05[1+w(a=0.5)]$ for large classes of DE models
\citep[e.g.][]{linder05}. Therefore, testing the precision with which
$\gamma$ can be measured should be regarded as a test of the precision
with which General Relativity (GR) can be verified on cosmological
scales. For instance, $\gamma\simeq 0.68$ corresponds to the linear
growth predicted by the popular DGP model of modified gravity
\citep{dgp00}. 

\begin{figure}
\hspace{-1.7truecm}
\includegraphics[width=0.47\textwidth,angle=270]{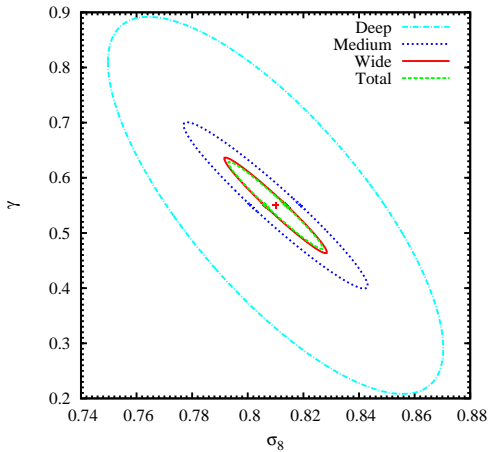}
\caption{Constraints on the power-spectrum normalization $\sigma_8$,
  and on the growth index $\gamma$ from the three WFXT surveys and
  from their combination.}
\label{fig:gamma}
\end{figure}

In order to test the constraints on the growth of perturbations
obtainable from the WFXT surveys, we freeze cosmic expansion to
$\Lambda$CDM, under the assumption that any modified gravity model
should be almost indistinguishable from $\Lambda$CDM at the background
level. Furthermore, we do not include any Dark Energy in our non-GR
models, following the idea that a modification of GR should be
alternative to DE to explain cosmic expansion. The results of our
analysis based on the WFXT surveys are shown in Figure
\ref{fig:gamma}. Here we report the expected constraints on the
$\gamma$--$\sigma_8$ plane, once we marginalize over the remaining
parameters. This plot confirms that WFXT would indeed provide very
useful constraints on possible deviations from the standard gravity,
based on the growth of structure as traced by the evolution of the 
cluster population.

\section{Synergies \& legacy value}
Addressing the outstanding questions outlined above will greatly
benefit from a coordinated multi-wavelength activity between WFXT,
future space missions and ground-based facilities (see also Rosati et
al. in this volume, for a more detailed description of the synergies
between WFXT and future instrumentation).

The identification and characterization of the galaxy populations
hosted by the $\sim 2\times 10^5$ clusters at $z>0.5$, unveiled by
WFXT, will be an essential process to obtain a comprehensive and
self-consistent picture of the cosmic cycle of baryons in their hot
and cold phase, by tracing the evolution of their underlying stellar
populations and star formation histories. Deep optical coverage of
large survey areas will be provided by the next generation of
wide-field ground-based facilities, currently under development and
scheduled for routine operations within the next few years, such as
Pan-Starrs\footnote{\tt http://pan-starrs.ifa.hawaii.edu/} and
LSST\footnote{\tt http://www.lsst.org/}.

The combination of WFXT with the ESA Euclid and the NASA JDEM
missions, currently under development (Euclid and JDEM) will provide
spectroscopic confirmation for a large fraction of $z>0.5$ clusters
identified by WFXT and a full characterization of member galaxies with
high resolution optical imaging. Such Dark Energy missions are also
designed to reconstruct the DM mass distribution via weak lensing
tomographic techniques. This will allow direct lensing mass
determination of thousands of massive clusters out to $z\sim 1$.
Their comparison with X-ray derived masses will yield the much
heralded cluster mass calibration and control of systematics for
cosmological applications.

The Atacama Cosmology Telescope (ACT) and the South Pole Telescope
(SPT) have recently opened a new era of Sunyaev-Zeldovich (SZ) cluster
search \citep{staniv09}. Next generation large single-dish mm
telescopes, such as the Caltech-Cornell Atacama Telescope\footnote{\tt
  http://www.submm.org/} (CCAT) will have enough sensitivity and
angular resolution to carry out large-area SZ surveys, providing at
the same time spatially resolved SZ imaging for moderately distant
massive clusters. Taking advantage of the different dependence of the
SZ and X-ray signals on gas density and temperature, their combination
will provide a reconstruction of temperature and mass profiles,
independent of X-ray spectroscopy
\citep[e.g.][]{ameglio09,golwala09}. This will offer further
independent means of calibrating mass measurements of clusters.

With its unprecedented grasp and angular resolution, WFXT will be an
outstanding source of interesting targets for follow-up studies of
galaxy clusters with facilities such as JWST, ALMA, ELT and future
X-ray observatories (i.e., IXO and Gen-X). For example, a combined
study of X-ray luminous proto-cluster regions with ALMA, will test
whether a phase of vigorous star formation (sub-mm bright galaxies)
coexist with a BH accretion phase.  Follow-up pointed observations
with IXO of extreme clusters identified by WFXT at $z\sim 2$ will
allow the study of metallicity and entropy structure of the pristine
ICM. In general, the synergy with next generation multi-wavelength
deep wide-area surveys and with high sensitivity instruments for
pointed observations will unleash the full potential of WFXT in
addressing a number of outstanding scientific questions related to
cosmological and astrophysical applications of galaxy clusters.

\begin{acknowledgements}
  This work has been supported by ASI--AAE Grant for Mission
  Feasibility Studies and by the INFN PD51 grant. The Authors would
  like to thank all the members of the WFXT team for a number of
  enlightening discussions.
\end{acknowledgements}

\bibliographystyle{aa}
\bibliography{master}
\end{document}